\documentclass[apj]{emulateapj}
\usepackage{epsfig}
\usepackage{amsmath}
\usepackage{natbib}
\usepackage{graphicx,subfigure}
\usepackage{calc}
\usepackage{verbatim}
\usepackage{tabularx}

\newcommand{\ztwo}{$z\sim2 \hspace{4pt}$}

\newcommand{\ls}{\hspace{2pt}}
\newcommand{\ha}{H$\alpha$\ls}				    
\newcommand{\nii}{[N{\small II}]\ls}                                       
\newcommand{\sii}{[S{\small II}]}                                       

\newcommand{\niiha}{[N{\small II}]/\ha}				    
\newcommand{\siiha}{[S{\small II}]/\ha}				    

\newcommand{\msun}{M$_{\odot}$} 			   
\newcommand{\sigsfrunits}{M$_{\odot} yr^{-1} kpc^{-2}$}
\newcommand{\mstar}{m$_{*}$} 				    
\newcommand{\mdotout}{$\dot{\rm M}_{\rm out}$}		

\newcommand{\siggas}{$\Sigma_{gas}$}
\newcommand{\sigsfr}{$\Sigma_{SFR}$}
\newcommand{\rhalf}{R$_{1/2}$}

\newcommand{\lcdm}{$\Lambda${\rm CDM}}		     

\shorttitle{Outflows from \ztwo galaxies}
\shortauthors{S. F. Newman et al.}

\begin{document}

\title{The SINS/zC-SINF survey of \ztwo galaxy kinematics: outflow properties$^{*}$}
\author{Sarah F. Newman$^{1,16}$, Reinhard Genzel$^{1,2,3}$, Natascha M. F\"orster-Schreiber\footnotemark[2], Kristen Shapiro Griffin\footnotemark[4], Chiara Mancini\footnotemark[7],  Simon J. Lilly\footnotemark[6], Alvio Renzini\footnotemark[7], Nicolas Bouch\'e$^{8,9}$, Andreas Burkert\footnotemark[10], Peter Buschkamp\footnotemark[2], C. Marcella Carollo\footnotemark[6], Giovanni Cresci\footnotemark[11], Ric Davies\footnotemark[2], Frank Eisenhauer\footnotemark[2], Shy Genel\footnotemark[5], Erin K. S. Hicks\footnotemark[12], Jaron Kurk\footnotemark[2], Dieter Lutz\footnotemark[2], Thorsten Naab\footnotemark[13], Yingjie Peng\footnotemark[6], Amiel Sternberg\footnotemark[14], Linda J. Tacconi\footnotemark[2], Daniela Vergani\footnotemark[15], Stijn Wuyts\footnotemark[2], and Gianni Zamorani\footnotemark[15]}

\footnotetext[*]{Based on observations at the Very Large Telescope (VLT) of the European Southern Observatory (ESO), Paranal, Chile (ESO program IDs 076.A-0527, 079.A-0341, 080.A-0330, 080.A-0339, 080.A-0635, 183.A-0781).}
\footnotetext[1]{Department of Astronomy, Campbell Hall, University of California, Berkeley, CA 94720, USA}
\footnotetext[2]{Max-Planck-Institut f\"ur extraterrestrische Physik (MPE), Giessenbachstr.1, D-85748 Garching, Germany}
\footnotetext[3]{Department of Physics, Le Conte Hall, University of California, Berkeley, CA 94720, USA}
\footnotetext[4]{Space Sciences Research Group, Northrop Grumman Aerospace Systems, Redondo Beach, CA 90278, USA}
\footnotetext[5]{Harvard-Smithsonian Center for Astrophysics, 60 Garden Street, Cambridge, MA 02138 USA}
\footnotetext[6]{Institute of Astronomy, Department of Physics, Eidgen\"ossische Technische Hochschule, ETH Z\"urich, CH-8093, Switzerland}
\footnotetext[7]{Osservatorio Astronomico di Padova, Vicolo dellÕOsservatorio 5, Padova, I-35122, Italy}
\footnotetext[8]{Universit\'e de Toulouse; UPS-OMP; IRAP; Toulouse, France}
\footnotetext[9]{CNRS; IRAP; 14, avenue Edouard Belin, F-31400 Toulouse, France}
\footnotetext[10]{Universit\"ats-Sternwarte Ludwig-Maximilians-Universit\"at (USM), Scheinerstr. 1, M\"unchen, D-81679, Germany}
\footnotetext[11]{ Istituto Nazionale di AstrofisicaÐOsservatorio Astronomico di Arcetri, Largo Enrico Fermi 5, I Ð 50125 Firenze, Italy}
\footnotetext[12]{Department of Astronomy, University of Washington, Box 351580, U.W., Seattle, WA 98195-1580, USA}
\footnotetext[13]{Max-Planck Institute for Astrophysics, Karl Schwarzschildstrasse 1, D-85748 Garching, Germany}
\footnotetext[15]{INAF Osservatorio Astronomico di Bologna, Via Ranzani 1, 40127 Bologna, Italy}
\footnotetext[14]{School of Physics and Astronomy, Tel Aviv University, Tel Aviv 69978, Israel}
\footnotetext[16]{email: sfnewman@berkeley.edu}


\begin{abstract}
Using SINFONI H$\alpha$, \nii and \sii \ls AO data of 27 \ztwo star-forming galaxies (SFGs) from the SINS and zC-SINF surveys, we explore the dependence of outflow strength (via the broad flux fraction) on various galaxy parameters. For galaxies that have evidence for strong outflows, we find that the broad emission is spatially extended to at least the half-light radius ($\sim$ a few kpc). Decomposition of the \sii \ls doublet into broad and narrow components suggests that this outflowing gas probably has a density of $\sim$10--100 cm$^{-3}$, less than that of the star forming gas (600 cm$^{-3}$). There is a strong correlation of the \ha broad flux fraction with the star formation surface density of the galaxy, with an apparent threshold for strong outflows occurring at 1 \msun yr$^{-1}$ kpc$^{-2}$. Above this threshold, we find that SFGs with log\mstar \textgreater 10 have similar or perhaps greater wind mass loading factors ($\eta$ = \mdotout/SFR) and faster outflow velocities than lower mass SFGs,  suggesting that the majority of outflowing gas at \ztwo may derive from high-mass SFGs. The mass loading factor is also correlated with the SFR, galaxy size and inclination, such that smaller, more star-forming and face-on galaxies launch more powerful outflows. We propose that the observed threshold for strong outflows and the observed mass loading of these winds can be explained by a simple model wherein Òbreak-outÓ of winds is governed by pressure balance in the disk. \\
\end{abstract}

\keywords{cosmology: observations --- galaxies: high-redshift --- galaxies: evolution --- infrared: galaxies }


\section{Introduction} 

Most high-z SFGs from rest-UV/optical samples show evidence for powerful galactic outflows, as indicated by UV absorption spectroscopy \citep{Pet+00,Shapley+03,Ste+10,Wei+09,Kor+12} and broad \ha emission-line profiles \citep{Sha+09,Gen+11,New+12}. This `star formation feedback' may be an essential ingredient in the evolution of high-z star forming galaxies, particularly between z $\sim$ 1--3, at the peak of the star formation rate density \citep{HopBea06}. However, little is yet known about how galaxy parameters determine the prevalence and strength of these outflows.

Current theoretical models suggest that outflows could be driven by energy or momentum feedback \citep[e.g.][]{Mur+05}. In the simple momentum driven wind scenario of \cite{OppDav06,OppDav08}, the mass loading parameter of the wind primarily depends on the circular velocity of the disk v$_{d}$, $\eta \propto$ v$_{d}^{-1} \propto$ M$_{baryon}^{-1/3}$, if the wind outflow velocity is near the escape velocity \citep{Mur+05}. \cite{Hop+12} have recently carried out high-resolution SPH simulations of isolated galaxies with different types of input feedback, including that from supernovae, stellar winds, expanding HII regions and radiation pressure and find an overall scaling of $\eta \propto$ v$_{d}^{-1} \Sigma_{gas}^{-1/2}$ and $\eta \sim$ 1 for the parameters of typical high-z massive SFGs and with energy-driven winds. The scaling with v$_{d}^{-1}$ is consistent with that found by \cite{OppDav06,OppDav08}, yet the dependence on \siggas$^{-1/2}$ is contrary to the findings of \cite{Che+10}, who found a positive correlation of Na D equivalent width (EW) with \sigsfr \ls from SDSS data of $\sim$100,000 galaxies, where Na D EW is used here as a proxy for the mass loading ($\eta$).

In this paper we analyze the outflow properties of 27 \ztwo SFGs, discussed in more detail in \cite{For+12} and based on new high-quality \ha emission line SINFONI/VLT integral field (IFU) spectroscopy with adaptive optics (AO) \citep{Eis+03,Bon+04}. We adopt a \lcdm$~$ cosmology with $\Omega_{m}$ =0.27, $\Omega_{b}$=0.046 and H$_{0}$=70 km/s/Mpc \citep{Kom+11}, and a \cite{Cha03} initial stellar mass function (IMF).
 
\section{Observations, Data Reduction and Analysis Techniques}

The galaxy sample discussed here (see Table 1) is described in \cite{Man+11,For+12}. Our 27 z$\sim$2--2.5 SFGs are drawn from the SINS and zC-SINF surveys of H$\alpha$+\nii integral field spectroscopy with SINFONI on the ESO VLT \citep{For+09,Man+11,For+12}. They were selected either from their U$_{n}$GR colors satisfying the `BX' criteria \citep{Ste+04,Ade+04,Erb+06,Law+09} or based on \textit{K} band imaging via the `BzK' criterion for 1.4 \textless z \textless 2.5 SFGs \citep{Dad+04}. They sample the z$\sim$2.2 star-forming `main sequence' for stellar masses between 10$^{9.5}$ to 10$^{11.5}$ \msun. Our sample includes 25 galaxies observed with AO-mode with sufficient SNR for analysis (0.05'' pixels, typical FWHM 0.2--0.25'') in addition to 2 galaxies observed in seeing limited mode that have very extended disks and thus are well resolved with 0.125'' pixels and FWHM$\sim$0.5''. The data are of high quality ($\sim$ 6h average integration time per galaxy, with a range from 2 to 22h, and a total integration time for the entire sample of 180h) and were reduced with our standard data reduction methods and analysis tools \citep{Sch+04,Dav+07,For+09,Man+11}.

\begin{table}
\caption{Galaxy Properties}
\scriptsize{
\renewcommand{\tabcolsep}{2pt}
\begin{tabular*}{0.5\textwidth}{l c c c c}
\\
\hline
\hline \\
\textbf{Galaxy ID} & \textbf{Reference} & \textbf{\mstar} & \textbf{SFR (H$\alpha$)} & \textbf{\sigsfr} \\
 & & (10$^{11}$\msun) & (\msun/yr) & (\sigsfrunits) \vspace{2pt} \\
 \hline \\
 Q1623-BX455 & 1,2 & 0.10 $\pm$ 0.031 & 63 $\pm$ 19 & 1.66 $\pm$ 0.52 \\
 Q1623-BX502 & 1,2 & 0.023 $\pm$ 0.0070 & 40 $\pm$ 12 & 2.86 $\pm$ 0.90 \\
 Q1623-BX543 & 1,2 & 0.094 $\pm$ 0.028 & 183 $\pm$ 55 & 2.38 $\pm$ 0.98 \\
 Q1623-BX599 & 1,2 & 0.57 $\pm$ 0.17 & 131 $\pm$ 39 & 3.95$\pm$ 1.63 \\
 SSA22a-MD41& 1,2,3 & 0.077 $\pm$ 0.010 & 131 $\pm$ 39 & 0.80 $\pm$ 0.24 \\
 Q2343-BX389 & 1,2 & 0.41 $\pm$ 0.080 & 196 $\pm$ 59 & 0.74 $\pm$ 0.22 \\
 Q2343-BX513 & 1,2 & 0.27 $\pm$ 0.081 & 28 $\pm$ 8 & 1.00 $\pm$ 0.36 \\
 Q2343-BX610 & 1,2 & 1.00 $\pm$ 0.27 & 212 $\pm$ 64 & 1.15 $\pm$ 0.35 \\
 Q2343-BX482 & 1,2 & 0.18 $\pm$ 0.080 & 121 $\pm$ 36 & 0.61 $\pm$ 0.19 \\
 GMASS-2303 & 4 & 0.072 $\pm$ 0.022 & 30 $\pm$ 9 & 0.52 $\pm$ 0.16 \\
 GMASS-2363 & 4 & 0.22 $\pm$ 0.065 & 50 $\pm$ 15 & 1.17 $\pm$ 0.37 \\
 ZC-401925 & 5,6,7 &  0.052 $\pm$ 0.016 & 13 $\pm$ 4 & 0.19 $\pm$ 0.060 \\
 ZC-404221 & 5,6,7 &  0.11 $\pm$ 0.050 & 60 $\pm$ 18 & 1.80 $\pm$ 0.57 \\
 ZC-405501 & 5,6,7 &  0.10 $\pm$ 0.030 & 68 $\pm$ 20 & 0.26 $\pm$ 0.10 \\
 ZC-405226 & 5,6,7 &  0.092 $\pm$ 0.028 & 82 $\pm$ 25 & 0.34 $\pm$ 0.17 \\
 ZC-406690 & 5,6,7 &  0.44 $\pm$ 0.13 & 480 $\pm$ 144 & 1.87 $\pm$ 0.57 \\
 ZC-407376 & 5,6,7 & 0.20 $\pm$ 0.060 & 283 $\pm$ 85 & 2.43 $\pm$ 0.77 \\
 ZC-407302 & 5,6,7 &  0.30 $\pm$ 0.090 & 260 $\pm$ 78 & 2.72 $\pm$ 0.83 \\
 ZC-409985 & 5,6,7 & 0.12 $\pm$ 0.036 & 40 $\pm$ 12 & 0.81 $\pm$ 0.25 \\
 ZC-410041 & 5,6,7 & 0.042 $\pm$ 0.013 & 105 $\pm$ 31 & 0.72 $\pm$ 0.22 \\
 ZC-410123 & 5,6,7 & 0.042 $\pm$ 0.013 & 40 $\pm$ 12 & 0.30 $\pm$ 0.090 \\
 ZC-411737 & 5,6,7 & 0.040 $\pm$ 0.015 & 60 $\pm$ 18 & 1.06 $\pm$ 0.33 \\
 ZC-412369 & 5,6,7 & 0.18 $\pm$ 0.050 & 182 $\pm$ 55 & 1.72 $\pm$ 0.60 \\
 ZC-413507 & 5,6,7 & 0.088 $\pm$ 0.025 & 92 $\pm$ 28 & 1.19 $\pm$ 0.43 \\
 ZC-413597 & 5,6,7 & 0.070 $\pm$ 0.020 & 86 $\pm$ 26 & 1.72 $\pm$ 0.55 \\
 ZC-415876 & 5,6,7 & 0.084 $\pm$ 0.022 & 42 $\pm$ 13 & 2.33 $\pm$ 0.80 \\
SA12-6339 & 5,6,7 & 0.26 $\pm$ 0.077 & 681 $\pm$ 204 & 17.34 $\pm$ 9.41 \\
\hline

\end{tabular*}}

References for parent sample. (1) \cite{Erb+06}, (2) \cite{Ste+04}, (3) \cite{Erb+03}, (4) \cite{Kur+09}, (5) \cite{Lil+07}, (6) \cite{Lil+09}, (7) \cite{McCra+10}.
\end{table}

Stellar masses and ages are derived from SED modeling in \cite{For+09,Man+11} assuming constant or exponentially declining star formation histories with \cite{BruCha03} tracks. The SFRs are derived from the SED modeling and from \ha (SFR=L$_{H\alpha}$/2.1x10$^{41}$ erg/s, \cite{Ken98}, corrected for a \cite{Cha03} IMF) with a \cite{Cal+00} reddening law with $A_{V,gas}=2.3 \times A_{V,SED}$. Half light radii (\rhalf) were computed from the cumulative \ha \ls flux profile determined after fitting a 2D exponential (Sersic n = 1) to the data. We calculate molecular gas masses (M$_{gas}$) and surface densities (\siggas = 0.5 $\times$ M$_{gas}$/($\pi R_{1/2}^{2}$)) from the H$\alpha$-derived SFRs using the molecular gas to star formation surface density relation with M$_{gas}$(\msun) = 5.8$\times 10^{8}$ x SFR (\msun yr$^{-1}$) \citep{Tac+12}. 
  
To isolate the broad emission for coadding, we remove large-scale rotational velocity gradients from the data cubes by aligning the spectral axis of each pixel, such that the \ha peak lies at the same velocity throughout the cube. This technique minimizes the width of the galaxy integrated spectrum due to large scale motions (e.g. rotation) in the narrow velocity component, and thus maximizes the contrast between the broad and narrow components. \cite{Sha+09} demonstrated that this method does not create artificial broad wings. From these velocity-shifted data cubes, we create spatially-integrated spectra for each galaxy by collapsing the cubes onto a single spectral axis \citep[see:][]{For+09,Sha+09}. Various stacks of spectra are generated by removing a fitted continuum from each spectrum, placing the spectra on a common rest-frame wavelength axis, weighting each spectrum by the signal to noise ratio (SNR) of \ha, and coadding the final spectra.
 
Broad emission has previously been detected in SINS/zC-SINF galaxies by \cite{Sha+09, Gen+11,New+12} and attributed to galactic outflows. We quantify the fraction of the emission line flux in this underlying broad component by simultaneously fitting two Gaussian functions to each of the H$\alpha$, [N{\small II}]$\lambda\lambda$6548,6584 and [S{\small II}]$\lambda\lambda$6716,6731 emission lines, setting the kinematics (velocity, line width) of all of the narrow Gaussian components to be equal to each other, and likewise setting the kinematics of the broad Gaussian components equal to each other. We allow the broad \ha flux to vary, but set the \niiha \ls ratio in the broad component to the \niiha \ls ratio in the narrow component (and likewise for \siiha). \cite{Gen+11} demonstrated that in the absence of a broad component, and after removal of large scale velocity shifts as described above, the remaining velocity profiles can be well fit by Gaussians. The `broad flux fraction' is defined as the ratio of the flux in the broad \ha emission line to the flux in the narrow \ha emission line.

We also generate position-velocity (pv) data by extracting the spectra along the major galaxy axis for each shifted cube with a slit wide enough to cover most of the minor axis emission. The data are then co-added and re-sampled onto a normalized spatial scale of 0.07xR$_{1/2}$ (FWHM $\sim$ 0.3xR$_{1/2}$) and spectral scale of 40 km/s/pixel ($\sim$90 km/s FWHM).
 
\section{Outflow Properties in \ztwo SFGs} 

We ascertain that the outflows are driven by the star formation activity and not by an AGN as we have excluded any known AGN from our sample, find that the broad emission is extended and not concentrated towards the kinematic or morphological center (see section 3.1), and measure \niiha ratios compatible with the expected metallicity based on the \ztwo mass-metallicity relation \citep{Erb+06} and too low to be consistent with the presence of an AGN. In two cases we trace the origin of the outflows to individual giant star forming clumps embedded in the disk \citep{Gen+11,New+12}. We note that \cite{Gen+11} suggest that the bright emission coming from a clump in one of our galaxies (ZC407302) could be due to AGN activity if this clump is an external minor merger, rather than a star forming clump formed within the disk. However, there is no other evidence that this galaxy contains an AGN.

\subsection{Outflows are Spatially Extended}

The broad emission (FWHM $\sim$450 km/s) is spatially extended over the half-light radii (\rhalf), which corresponds to a few kpc in the small galaxies and up to 6 kpc in the larger disks. This is demonstrated in the bottom panel of Figure 1, where we show a co-added pv diagram of the line emission (large-scale velocity gradients removed) along the morphological major axis for the galaxies. Broad emission is detected at least out to \rhalf, and its width between 0.5 x R$_{1/2}$ and R$_{1/2}$ (FWHM = 423 $\pm$ 19 km/s) is nearly as broad as within 0.5 x R$_{1/2}$ (FWHM = 475  $\pm$ 12 km/s). The upper panel of Figure 1 shows that the emission in the outer regions of the galaxies is almost as broad as in the inner regions, and implies a constant or decreasing mass-loading with increasing galacto-centric radius.

The 10 galaxies used in this coadd (BX455, BX513, BX543, BX599, SA12-6339, ZC404221, ZC407306, ZC407302, ZC412369, ZC415876) are selected such that they have noticeable broad line wings without the need for stacking and an absence of OH residuals near H$\alpha$. All but one have \rhalf \textless 3 kpc and all have \sigsfr \textgreater 1 \sigsfrunits. However, their stellar masses and SFRs span the ranges seen in our full sample. We note that although most of the galaxies from this stack are small (with \rhalf \ls \textless 3 kpc), we also observe broad emission from stacks of larger galaxies that also have \sigsfr \textgreater 1 \sigsfrunits \ls (see next section).

\begin{figure*}[tb]
\centerline{
\includegraphics[width=5in]{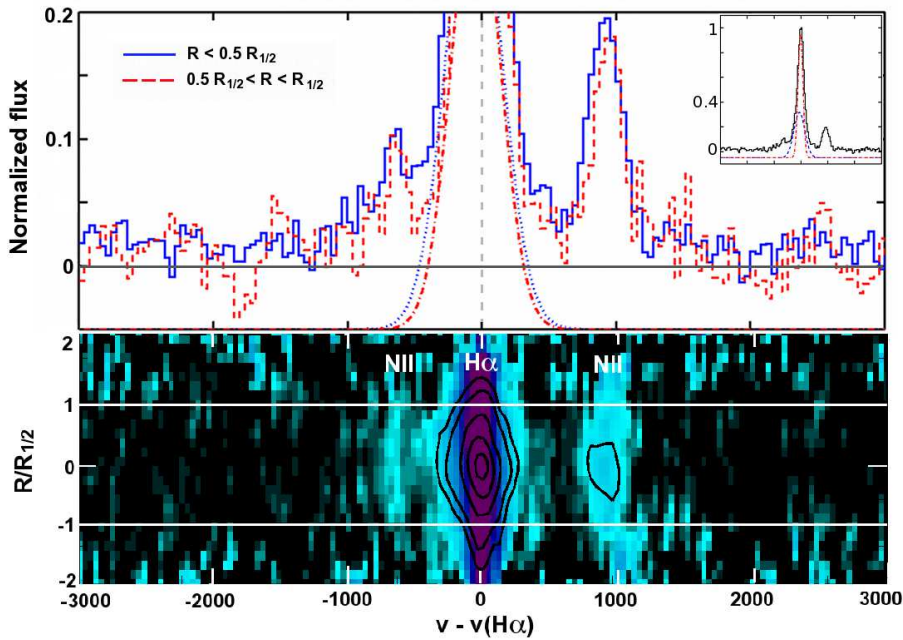}}
\caption{Lower panel: Co-added (pv) diagram for the 10 sources with the best evidence for broad emission, with the vertical (spatial) axis normalized to \rhalf. The black contours show the unnormalized stack, while the colors have been normalized according to the peak \ha flux for each spatial row. The broad emission in light blue (\textgreater 200--300 km/s from the systemic velocity) is spatially extended (vertical axis) out to at least \rhalf. Upper panel: Spectra from the inner (R\textless 0.5 x \rhalf, blue, solid line) and outer (0.5 x \rhalf \textless R\textless \rhalf, red, dashed line) regions of the galaxies, with the broad fitted components shown by the blue dotted line (inner) and red dash-dotted line (outer). The inner and outer region spectra have F$_{broad}$/F$_{narrow}$ = 1.31 $\pm$ 0.075 and 1.13 $\pm$ 0.12 with broad to narrow component velocity shifts of -41 $\pm$ 5 km/s and -33 $\pm$ 9 km/s, respectively. The inset shows the entire \ha line for the inner spectrum (black) with the broad component (blue, dashed) and narrow component (red, dash-dot) overplotted.}
\end{figure*}

\subsection{The Broad Flux Fraction is Strongly Dependent on the SF Surface Density}

We explore the dependence of the broad flux fraction on galaxy properties by dividing the entire sample into two bins each (low/high) in star formation rate, stellar mass, size, inclination and star formation surface density. We then co-added the spectra in each of the bins and computed the ratio of the broad to narrow \ha emission. The results are summarized in Table 2.

\begin{table*}
\caption{Broad flux fraction for high and low bins of different galaxy properties}
\scriptsize{
\renewcommand{\tabcolsep}{8pt}
\begin{tabular*}{1.0\textwidth}{l c c c c c c c c}
\\
\textbf{Property} & \multicolumn{2}{c}{F$_{broad}$/$F_{narrow}$ (\ha)} & \multicolumn{2}{c}{FWHM$_{broad}$ (km/s)} & \multicolumn{2}{c}{$\Delta \rm v_{broad-narrow}$ (km/s)} & Significance & Dividing \\
 & High bin & Low bin & High bin & Low bin & High bin & Low bin & of difference & value\\
\hline
\hline
\\
\textbf{SFR} & 0.65 $\pm$ 0.074 & 0.50 $\pm$ 0.041 & 510 $\pm$ 12 & 423 $\pm$ 47 & -23 $\pm$ 4 & -21 $\pm$ 8 & 2$\sigma$ & 100 \msun /yr \\
\textbf{\rhalf} & 0.50 $\pm$ 0.054 & 0.76 $\pm$ 0.082 & 503 $\pm$ 15 & 432 $\pm$ 19 & -15 $\pm$ 6 & -35 $\pm$ 6 & 3$\sigma$ & 3 kpc \\
\bf{\sigsfr} & 0.77 $\pm$ 0.027 & 0.16 $\pm$ 0.030 & 500 $\pm$ 16 & 423 $\pm$ 75 & -27 $\pm$ 4 & -43 $\pm$ 23 & 20$\sigma$ & 1 \sigsfrunits \\
\textbf{inclination} & 0.47 $\pm$ 0.055 & 0.77 $\pm$ 0.091 & 510 $\pm$ 44  & 514 $\pm$ 12  & -39 $\pm$ 14 & -32 $\pm$ 5 & 3$\sigma$ & 49--55$^{o}$\\ 
 \textbf{m$_{*} \ls ^{1}$} & 0.63 $\pm$ 0.056 & 0.41 $\pm$ 0.042 & 528 $\pm$ 13 & 423 $\pm$ 45 & -32 $\pm$ 5 & -20 $\pm$ 8 & 4$\sigma$ & 1x10$^{10}$ \msun \\
 \textbf{m$_{*} \ls^{2}$} & 0.23 $\pm$ 0.21 & 0.12 $\pm$ 0.040 & 423 $\pm$ 66 & 423 $\pm$ 80 & -28 $\pm$ 18 & -48 $\pm$ 42 & 0.5$\sigma$ & '' \\
  \textbf{m$_{*} \ls^{3}$} & 0.85 $\pm$ 0.097 & 0.71 $\pm$ 0.10& 520 $\pm$ 11 & 428 $\pm$ 30 & -25 $\pm$ 4 & -13 $\pm$ 7 & 1.5$\sigma$ & '' \\
 \hline
 \\
\end{tabular*}}

The high/low bins for SFR, \mstar, \sigsfr, \rhalf, and inclination are divided above and below the   value(s) shown in column 9 and have roughly an equal number of galaxies in each bin. The spectrum from each stack was fit constraining the narrow FWHM to between 190 and 210 km/s, so that the broad component was not fit as a very broad narrow component, and allowing the relative velocities of the two components and the broad line width (FWHM) to vary. The best fit broad FWHM and the relative velocities for each bin are shown in columns 4--7. For the \mstar \ls bins, (1) is for all galaxies, (2) is for galaxies with \sigsfr \textless 1 \sigsfrunits, and (3) is for galaxies with \sigsfr \textgreater 1 \sigsfrunits. \\
\end{table*}

The star formation surface density has the largest effect on the broad flux fraction, such that galaxies with \sigsfr \textgreater 1 \sigsfrunits \ls drive the strongest outflows, which confirms and strengthens a similar finding in \cite{Gen+11}. This result is also consistent with what was observed in z $\sim$ 0 SFGs by \cite{Che+10}, where they found a strong correlation of Na D absorption equivalent width (from SDSS data of $\sim$ 100,000 galaxies) with \sigsfr.

In Figure 2, we see the dependence of the broad \ha flux fraction on \sigsfr \ls from 5 \sigsfr-binned points as well as three massive star-forming clumps. The trend seen in Figure 2 can be well described as a `threshold' for outflows. All of the data points with \sigsfr \ls \textgreater 1 \sigsfrunits \ls have F$_{broad}$/F$_{narrow} >$ 0.7, while all those below this threshold have F$_{broad}$/F$_{narrow} \leq$ 0.25. We also compare the spectra and broad component fits for stacks of galaxies above and below the \sigsfr \ls threshold in the lower panels of Figure 2. For these high \sigsfr \ls and low \sigsfr \ls stacks, the two-component fits yield a reduced $\chi^{2}$ of 1.15 and 1.55, respectively, while an imposed single-component fit yields reduced $\chi^{2}$ values of 16.9 and 3.27, indicating that a two-component fit is much better in the stack with a higher F$_{broad}$/F$_{narrow}$ ratio, and thus stronger wind signature, and somewhat better in the stack with a smaller outflow signature.

The broad flux fraction also varies (to a lesser extent) with \mstar, SFR, \rhalf \ls and inclination such that higher F$_{broad}$/F$_{narrow}$ values and thus stronger outflows are observed for massive, high SFR, compact, face-on galaxies. We further explore the \mstar \ls trend by dividing each of the high and low \sigsfr \ls bins into high and low \mstar \ls bins. Unsurprisingly, we find that both of the \mstar \ls bins below this threshold show negligible evidence for broad emission, and both high \sigsfr \ls bins show strong outflows. Above the \sigsfr \ls threshold, the broad flux fraction is similar or perhaps somewhat larger in the high \mstar \ls bin. The FWHM of the broad line is larger for the high \mstar \ls bin (520 $\pm$ 11 km/s vs. 428 $\pm$ 30 km/s) and more blueshifted (-25 $\pm$ 4 km/s vs. -13 $\pm$ 7 km/s). This increase is smaller than a `virial' scaling, v$_{out} \leq$ v$_{esc} \sim$ v$_{c} \sim$ \mstar$^{0.33}$ \citep{Mur+05,Mar05}, which would imply a factor of $\sim$1.8 in v$_{out}$ between the two mass bins (log\mstar = 9.85 and 10.61).

The SFR trend is linked to that of \mstar, due to the strong correlation of \mstar \ls and SFR for `main-sequence' SFGs at all redshifts. The \rhalf \ls trend is likely due to the higher \sigsfr \ls values for the more compact galaxies. The trend with inclination supports a bipolar outflow emerging perpendicularly from the plane of the star-forming disk, consistent with the results of \citet[at z $\sim$ 1]{Kor+12}, \citet[at z $\sim$ 0]{Che+10}, \cite{Bou+12} and \cite{Bor+11} (but see \cite{Law+12}).

\begin{figure}[tb]
\centerline{
\includegraphics[width=3.5in]{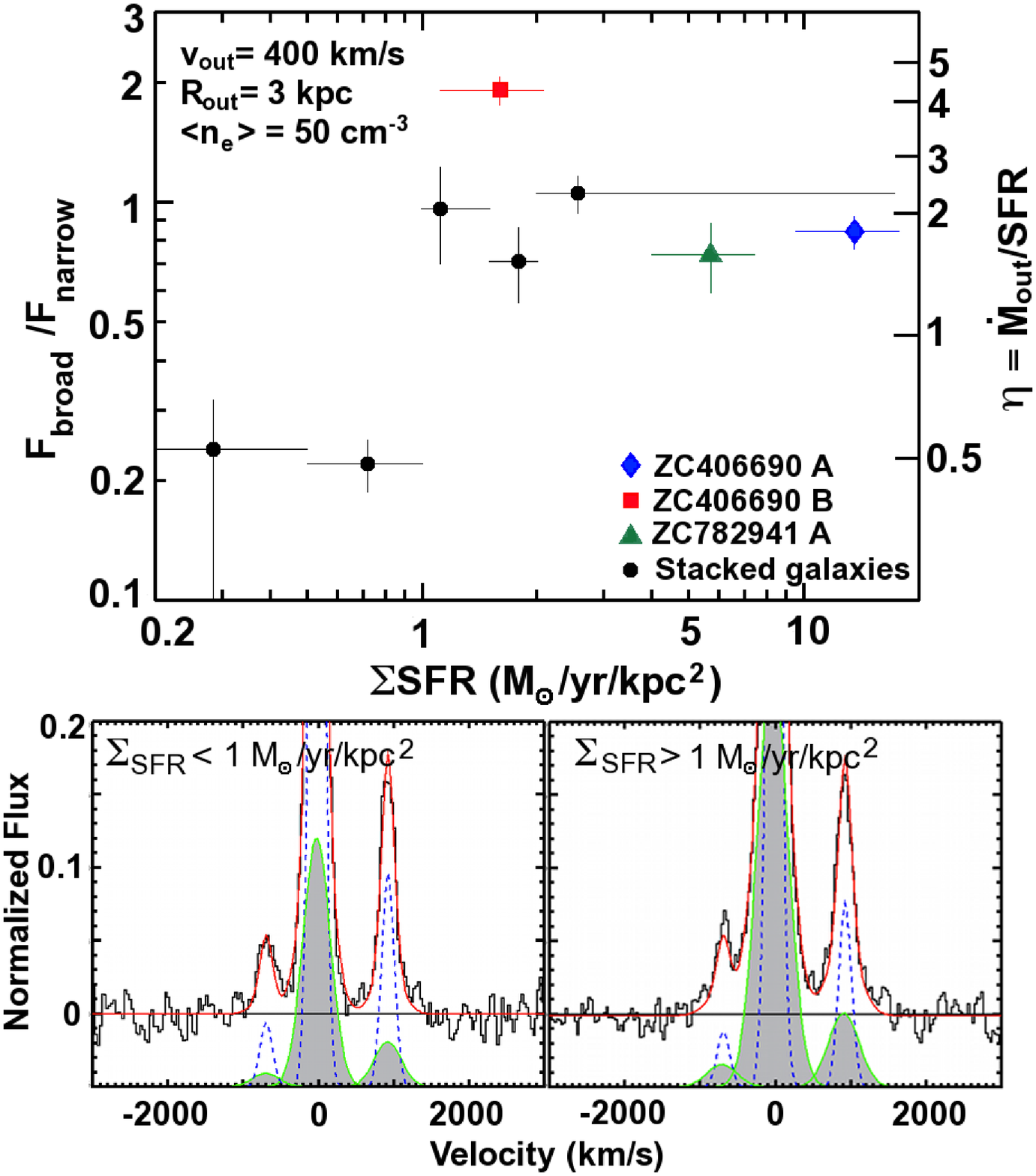}}
\caption{Dependence of the broad to narrow \ha flux ratio on \sigsfr. Above: Stacks in 5 star formation surface density bins with 5--8 galaxies each. The right vertical axis shows the corresponding mass loading with the model explained in the text and wind parameters listed in the figure. The black points are the stacked galaxies and the colored symbols are individual star-forming clumps. The horizontal error bars for the binned points represent the range in \sigsfr \ls for the galaxies in each bin. The vertical error bars represent the rms of the range in F$_{broad}$/F$_{narrow}$ values achieved by varying the stacks (for the binned points) and the measurement error (for the clumps). Below: Co-added \ha+\nii spectra of low and high \sigsfr \ls bins with the data in black, the best fit shown in red, the narrow components shown in blue, and the broad components shown in green with grey shading. 
}
\end{figure}

\subsection{Local Electron Density of the Outflow}

We estimate the ratio of the \sii \ls doublet in the broad and narrow components for a stack of the galaxy spectra, in order to constrain the star-forming gas and wind densities. The 14 galaxies used in this stack are selected such that they do not have strong OH sky features close to the location of the \sii \ls lines and have noticeable broad \ha components. We follow the fitting method described earlier in the text, except here we allow the amplitudes of the broad and narrow components in all nebular lines to vary, as opposed to, for instance, setting \niiha \ls (broad) equal to \niiha \ls (narrow). The reduced $\chi^{2}$ of the fit in the region of the \ha line is 1.71 and in the region of the \sii \ls lines is 0.97, while an imposed one-component fit yields reduced $\chi^{2}$ values of 26.1 and 1.24, respectively, highlighting the challenges of attempting such a measurement.

For the narrow component, we find F(\sii$\lambda$6716)/F(\sii$\lambda$6731) = 0.99 $\pm$ 0.27, and for the broad component, we find that the ratio = 1.43 $\pm$ 0.40, corresponding to electron densities of 600 (+1000/-450) and 10 (+590/-10) cm$^{-3}$, respectively \citep{Ost89}. The errors come from the fit uncertainties. Although our estimate is uncertain, the inferred local gas density in the outflow could be more than an order of magnitude less than the density of the star-forming regions (traced by the narrow component), consistent with a diffuse outflow. In \cite{New+12}, we assumed a mass outflow density of 100 cm$^{-3}$, based on two different outflow geometries and estimates of outflow density from local starburst galaxies \citep[see e.g.][]{HecArmMil90}. Thus we assume that these two values are upper and lower limits and take an average, resulting in a wind density of 50 cm$^{-3}$. Figure 3 shows the co-added spectrum along with the best fit.

A comparison of the narrow and broad \niiha \ls ratios (0.17 $\pm$ 0.017 and 0.31 $\pm$ 0.028, respectively) with those of \cite{New+12} in the clump and wind regions of  ZC406690 shows that the narrow \niiha \ls ratio is quite similar to the clump value and the broad ratio is very similar to that of the wind regions, which are affected by shocks. Therefore, not only does the broad \sii \ls ratio tell us about the density in the wind, but the broad \niiha \ls value provides further evidence that the broad component derives from outflows.

\begin{figure*}[tb]
\centerline{
\includegraphics[width=5in]{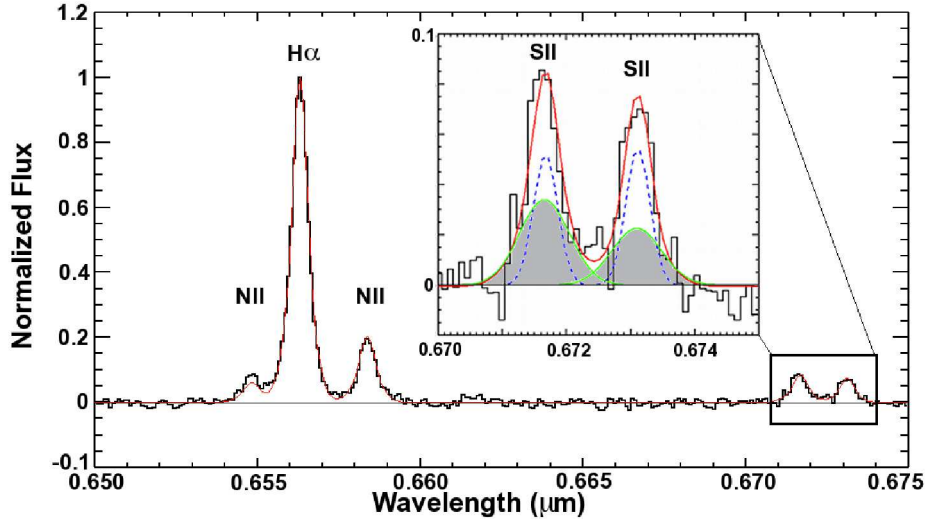}}
\caption{Co-added spectrum of 14 SFGs with [SII] detections that are free of OH contamination. The black line is the data, the red line is the best fit, and the green lines with grey shading (with FWHM = 407 km/s) and blue dashed lines (with FWHM = 197 km/s) (seen in the inset) are the broad and narrow Gaussian fits, respectively. 
}
\end{figure*}

\subsection{Mass-loading of High-z Galactic Outflows}

We assume the simplified outflow model of \cite{Gen+11} and \cite{New+12} for a warm ionized outflow with a radially constant outflow velocity and mass loss rate, to convert the \textit{F$_{broad}/F_{narrow}$} ratio in Figure 2 into a mass-loading factor $\eta \equiv$ \mdotout/SFR,

\begin{equation}
\dot{\rm M}_{\rm out} = M_{w} \times \frac{v_{out}}{R_{out}} = \frac{1.36 m_{H}}{\gamma_{H\alpha}n_{e}}\left(L_{H\alpha} \times \frac{F_{broad}}{F_{total}} \right) \frac{v_{out}}{R_{out}}
\end{equation}

\noindent where \mdotout \ls is the mass outflow rate, M$_{w}$ is the instantaneous mass in the outflow, v$_{out}$ is the average velocity of the outflow, R$_{out}$ is the radial extent, m$_{H}$ is the atomic mass, $\gamma_{H\alpha}$ is the \ha emissivity at T$_{e}$ = 10$^{4}$ K ($\gamma_{H\alpha} = 3.56 \times 10^{-25}$ erg cm$^{3}$ s$^{-1}$), \textit{n$_{e}$} is the local electron density in the outflow, \textit{L$_{H\alpha}$} is the total extinction-corrected \ha luminosity, and \textit{F$_{broad}$/F$_{total}$} is the fraction of the total flux in the broad component. We assume v$_{out}$ =  400 km/s (see section 4) and R$_{out}$ = 3 kpc. The latter is based on the spatial offset of outflowing gas from a massive star-forming clump described in \cite{New+12}, and from the finding in section 3.1 that the broad emission is extended to several kpc from the galaxy center. The emission line based estimate is independent of the collimation of the outflow. We adopt \textit{n$_{e}$} = 50 cm$^{-3}$ as the average \textit{local} electron density, derived in the previous section.

The inferred mass loading factors corresponding to the broad flux fraction are shown on the right-most y-axis in Figure 2. The mass-loading factors for galaxies below the \sigsfr \ls threshold are $\sim$ 0.5 and above the threshold are $\sim$ 2, albeit with an overall absolute uncertainty of at least a factor of 3. Despite the uncertainties, a mass loading factor of 2 is consistent with observations of both local starbursts \citep{HecArmMil90,Vei+05,Che+10} and high-z SFGs \citep{Pet+00,Wei+09,Ste+10,Gen+11,Bou+12}, as well as theoretical predictions \citep{Mur+05,Dave+11,Hop+12}.

\section{Why Do the Outflows Depend so Strongly on \sigsfr?}

We propose that the strong dependence of $\eta$ on \sigsfr \ls discussed in section 3.4 is mainly caused by the threshold that governs when star formation feedback can break out of the dense gas layer in the disk. 

Following \cite{OstShe11} (see Equations 1 and 7), in a baryon dominated galactic disk in pressure equilibrium, the weight of the disk balances the pressure generated from star-formation feedback in the form of supernovae, stellar winds, HII regions, cosmic rays and radiation from the star-forming layer. If that pressure exceeds the weight of the disk, then a momentum-driven outflow is launched perpendicular to the galactic plane, with a threshold \sigsfr \ls (\sigsfrunits) of,

\begin{equation}
\Sigma_{SFR,th} = \frac{\pi G f_{g}}{2 (P_{tot}/m_{*})} \Sigma_{d}^{2} = 0.9 \times \frac{f_{g,0.5} \times \Sigma_{d,500}^{2}}{(P_{tot}/m_{*})_{1000}}
\end{equation}

\noindent where G is the gravitational constant, (P$_{tot}$/m$_{*}$) is the characteristic total momentum injection per mass, $\Sigma_{d}$ is the disk surface density, and f$_{g}$ is the gas fraction, with fiducial values of 1000 km/s \citep{OstShe11,Mur+05}, 500 \msun  pc$^{-2}$ \citep{Erb+06,For+09} and 0.5 \citep{Tac+10,Dad+10}, respectively. Here, we have equated the weight of the gas above the disk \citep[see:][Equation 1, with an added dependency on f$_{gas}$ since we only care about infalling gas]{OstShe11} with the pressure from star formation feedback \citep[see:][Equation 7]{OstShe11}. This is similar to the ÔEddington limitÕ of momentum driven winds \citep{Mur+11}, which gives $\Sigma_{SFR,threshold} \propto v_{c}^{5/2} R^{-2}$, where v$_{c}$ is the circular velocity.

The wind has a total momentum outflow rate,

\begin{equation}
\dot{\rm M}_{\rm w} v_{\infty} \leq \left(\frac{P_{tot}}{m_{*}} \right) SFR
\end{equation}

\noindent where v$_{\infty}$ is the outflow velocity at large distances from the mid-plane, implying a mass-loading factor,

\begin{equation}
\eta = \frac{\dot{\rm M}_{\rm w}}{SFR} \leq \frac{(P_{tot}/m_{*})}{v_{\infty}} = 2.5 \times \frac{(P_{tot}/m_{*})_{1000}}{v_{\infty,400}} 
\end{equation}

\noindent with v$_{\infty,400}$ in units of 400 km/s. This velocity is motivated by the observed outflow velocities in both low-z and high-z SFGs \citep{Pet+00,Mar05,Vei+05,Wei+09,Ste+10,Gen+11} as well as from this work. The adopted value of the characteristic momentum injection is chosen assuming $P_{tot}/m_{*}$ from radiation pressure $\sim$ 200--300 km/s \citep{OstShe11} and that the direct momentum injection contributions of supernovae, stellar winds, HII regions and radiation pressure are all roughly comparable \citep{Mur+05,Mur+10}, giving a total ($P_{tot}/m_{*}$) of 1000 km/s. Here, we have assumed that energy-driven winds are unimportant and that the bulk of the ram pressure from SNe does not play a strong role until the gas has been lifted out of the disk \citep[see:][]{Mur+11}.

This simple model predicts that for constant $\eta$ the ability of an outflow to break out of the disk strongly depends on the star formation surface density, and this threshold value scales linearly with gas fraction and as the square of the disk surface density. Second, the mass loading of galactic winds above breakout and the threshold star formation surface density for breakout both depend on ($P_{tot}/m_{*}$). Thus, if one of these quantities can be determined for a galaxy of known properties, then the other can be predicted. The model also predicts that it is harder to launch a wind from a more massive galaxy (with larger v$_{c}^{4}$/R$^{2}$). This final prediction is also acquired by writing the threshold in terms of \mstar \ls by assuming (1) the \rhalf-\mstar \ls relation of \cite{Ich+12} (\rhalf $\sim$ \mstar$^{0.14}$), (2) the SFR-\mstar \ls relation for normal SFGs (SFR = \mstar$^{0.7}$ x ((1+z)/3.2)$^{2.6}$) \citep{Elb+07,Noe+07} and (3) a constant gas depletion timescale (t$_{depl}$ = M$_{g}$/SFR $\sim$ 6.4 x 10$^{8}$ yr, \cite{Tac+12}). We note that the \cite{Ich+12} relation was derived using K band continuum emission, while our \rhalf \ls are derived using \ha, so there may be some inconsistency. This gives an \mstar \ls threshold of $\sim$ few x 10$^{10}$ \msun, such that it is easier for outflows to `break-out' from galaxies below this threshold. 

The first two predictions are met by the available data. For z$\sim$0 normal SFGs with f$_{g}\sim$0.07 and $\Sigma_{d}\sim$500 \msun pc$^{-2}$ \citep{Cat+10}, Equation 4 suggests a critical break-out star formation surface density of $\sim$0.1 \sigsfrunits. And indeed, winds are only detected in compact star forming dwarfs and/or starbursts above $\sim$0.1 \sigsfrunits \citep{HecArmMil90,Vei+05,Che+10}, see also \cite{Mur+11}. Mass loading factors for local galaxies with winds are estimated to be around 1 \citep{Mar+12}.

For the typical high-z SFGs presented in this paper (f$_{g}\sim$ 0.5, $\Sigma_{d}\sim$ 500--1000 \msun pc$^{-2}$), Equation 2 suggests a breakout star formation surface density near or slightly above 1 \sigsfrunits, in remarkable agreement with our findings in section 3.2 and Figure 2. Above this threshold, Equation 4 predicts a mass loading of $\sim$2.5, again in good agreement with our observations. We caution that our determination of the mass loading factor in the previous section is dependent on a simplified model with fairly uncertain parameters, yet the agreement is encouraging.

A positive correlation of $\eta$ with \sigsfr \ls (which was also observed by \cite{Che+10}) is possible evidence against the scenario in which winds are launched by the energy of hot supernovae bubbles, as in this case the shorter cooling time of the dense gas suggests $\eta \propto \Sigma_{SFR}^{-1/2}$ \citep{Hop+12}, and could support the momentum-driven wind model.

However, the third prediction of the simple model (massive galaxies are less efficient at driving winds) does not appear to be met by our data. For the SFGs in the `wind regime' (\sigsfr \ls $>$ 1 \sigsfrunits) we do not see any significant variation of F$_{broad}$/F$_{narrow}$ as a function of v$_{c}$, which is expected theoretically for momentum driven winds \citep[$\eta \propto v_{c}^{-1}$,][]{Mur+05,OppDav06,OppDav08,Hop+12}. 

This finding as well as the approximate \mstar \ls independence of the observed F$_{broad}$/F$_{narrow}$ ratio in our data surprisingly suggest that massive SFGs have mass loading factors that are similar to or higher than those of low mass SFGs. If so, the volume averaged outflow rate and metal enrichment of the circum-galactic/inter-galactic medium at \ztwo may be dominated by massive galaxies just around the Schechter mass \citep[logM$_{*}$ \ls $\sim$ 10.65,][]{Pen+10}. Indeed, with a slope $\alpha$ of the main-sequence of SFGs (SFR $\propto m_{*}^{\alpha}$) $\sim$ 0.7 or higher, the SFR (and thus the outflow rate, with constant $\eta$) increases faster with \mstar \ls than the volume density of SFGs decreases ($\Phi$(\mstar) $\sim$ \mstar$^{-0.5}$, \cite{Pen+10}). We note that a massive galaxy dominance in the metal enrichment of the intracluster medium (ICM) is demanded by the fact that in clusters $\sim$ 2/3 of the metal mass is contained in the ICM whereas only $\sim$ 1/3 is still locked into stars and galaxies \citep{Ren97}.

Further, the \ztwo \mstar \ls dependence of the mass-metallicity relation \citep[Z $\sim$ \mstar$^{0.24}$,][]{Erb+06} may be primarily driven by the larger (diluting) gas fractions in low \mstar \ls SFGs, besides these galaxies driving winds more effectively (M$_{g}$/\mstar = t$_{depl}$ x SFR/\mstar $\sim$ \mstar$^{-0.3}$). In addition, if high-\mstar \ls galaxies are indeed ejecting this much mass, we may have uncovered a mechanism contributing to the quenching of star formation near the Schechter mass \citep{Pen+10}. 

\acknowledgements
We would like to thank the anonymous referee for a very thoughtful and useful review. We are grateful to Jerry Ostriker for a very valuable discussion on the wind breakout. SFN is supported by an NSF grfp grant. CM, AR, GZ and DV acknowledge partial support by the ASI grant ``COFIS-Analisi Dati'' and by the INAF grants ``PRIN-2008'' and ``PRIN-2010''.


\nocite{Hop+12}
\nocite{Che+10}
\nocite{Gen+11}
\nocite{Kor+12}
\nocite{Tac+10}
\nocite{Erb+06}
\nocite{Dad+10}
\nocite{New+12}
\nocite{Mur+05}
\nocite{OppDav06}
\nocite{OppDav08}
\nocite{HecArmMil90}
\nocite{Vei+05}
\nocite{OstShe11}
\nocite{Pet+00}
\nocite{Mar05}
\nocite{Wei+09}
\nocite{Ste+10}
\nocite{Bou+12}
\nocite{Dav+11}
\nocite{Dave+11}
\nocite{Cec+01}
\nocite{Vei+94}
\nocite{Wes+07}
\nocite{Law+12}
\nocite{Sha+09}
\nocite{For+09}
\nocite{Man+11}
\nocite{Cal+00}
\nocite{BruCha03}
\nocite{Sch+04}
\nocite{Dav+07}
\nocite{Kom+11}
\nocite{Cha03}
\nocite{Eis+03}
\nocite{Bon+04}
\nocite{Shapley+03}
\nocite{Ken98}
\nocite{Bor+11}
\nocite{Ost89}
\nocite{Mar05}
\nocite{Tre+04}
\nocite{Ich+12}
\nocite{Pen+10}
\nocite{Mur+11}
\nocite{Ren97}
\nocite{For+12}
\nocite{Tac+12}
\nocite{Elb+07}
\nocite{Noe+07}
\nocite{Cat+10}
\nocite{Mar+12}
\nocite{Dad+04}
\nocite{Ste+04}
\nocite{Ade+04}
\nocite{Law+09}
\nocite{HopBea06}
\nocite{Mur+10}
\nocite{Erb+03}
\nocite{Lil+07}
\nocite{Lil+09}
\nocite{McCra+10}
\nocite{Kur+09}

\end{document}